# Can we rely on smartphone applications?

Sonia Meskini[1], Ali Bou Nassif [2] and Luiz Fernando Capretz [3]

[1] Prophix Software, Mississauga, Ontario, L5B 3J1, Canada
`sonya.meskini@gmail.com`
[2] Department of Electrical & Computer Engineering, University of Sharjah, 27272, UAE
`anassif@sharjah.ac.ae`
[3] Department of Electrical and Computer Engineering, Western University, London, Ontario, N6A 5B9, Canada
`lcapretz@uwo.ca`

**Abstract.** Smartphones are becoming necessary tools in the daily lives of millions of users who rely on these devices and their applications. There are thousands of applications for smartphone devices such as the iPhone, Blackberry, and Android, thus their reliability has become paramount for their users. This work aims to answer two related questions: (1) Can we assess the reliability of mobile applications by using the traditional reliability models? (2) Can we model adequately the failure data collected from many users? Firstly, it has been proved that the three most used software reliability models have fallen short of the mark when applied to smartphone applications; their failures were traced back to specific features of mobile applications. Secondly, it has been demonstrated that the Weibull and Gamma distribution models can adequately fit the observed failure data, thus providing better means to predict the reliability of smartphone applications.

## 1 Introduction

Smartphones are now so useful that many people prefer them over desktop or laptop computers. Hundreds of applications, usually suited to desktop or laptop computers, have been adapted to and carried out by these smartphones. The high usage and trust placed in these devices and their applications make their reliability a critically important goal to achieve [1]. Thus, owing to their highly integrated software, smartphones are far more advanced devices and their functionalities far exceed those of the classic mobile phones. Therefore, increased attention is now being paid to the reliability and security of these devices. Software Reliability Growth Models (SRGMs) are among the tools that deal with the reliability of software applications; they have been constructed and successfully applied to desktop (classic/standard) applications. In recent work [2], we thoroughly investigated the applicability of these SRGMs to the mobile area. We applied three of the most used SRGMs to the collected failure data of three smartphone applications; our main conclusion was that none of the selected models was able to account for the observed failure data satisfactorily. Basically, we addressed the following research questions:

(1) How do the existing successful reliability models, used to assess the desktop/laptop applications, perform when applied to the mobile area?
(2) What are the best non-linear distributions that fit smartphone application failure data?
(3) What useful information can be gained from this approach?

The rest of the paper is organized as follows: in Section 2 we provide a short list of the existing models that we will use, we describe our dataset collection, and we test the applicability of existing software reliability models. Finally, in Section 3, we carry out an analysis of the failure data with model distributions followed by a discussion in Section 4. We present our conclusions in Section 5 and outline future work possibilities.

## 2 SRGMs Applied to Smartphone Applications

The SRGMs used later in our experiments are: the NHPP – Crow – AMSAA model (also termed the NHPP-Power Law model), the Musa-Basic execution time model (or the exponential model), and the Musa-Okumoto model (or the Logarithmic Poisson model). The applications that have been chosen are Skype, Vtok, and a private Windows phone application. The relevant equations of these models are given in [3].

We present the procedure devised to collect the failure data for each application followed by the results of the application of the chosen SRGM to failure data for each application, and, finally, an analysis of the observed results.

### 2.1 Datasets and experiments

We used Apple devices (iPhone, iPad and iPod Touch) crash files as well as a Windows Phone crash file as our "experimental" data. These crash files are not public, but are confidential.
The reliability demonstration of smartphone applications was carried out through traditional testing, failure data collection, and the application of the most used SRGMs for standard applications to observe and check the adequacy of these models in the mobile area.

The first iPhone application studied was Skype, which had been tested and used for one year (from November 1, 2011 to November 11, 2012). Hence, the data has been collected during this year with some missing values due to the occasional non-use of the application. We were, however, able to collect 39 data points for the Skype application.

The second application studied was Vtok (an application for Google talk). This application was used continuously every day for two months (from September 19, 2012 to November 25, 2012). Hence, we were able to collect failures every day (81 data points).

During these periods, both the Skype and the Vtok applications were upgraded when new versions were released.

On the other hand, the Windows phone application was used and tested continuously for six months (from March 2012 to August 2012) by different users located in different parts of the world (more than 100 users).

We used two Software Reliability tools for this application to double check the results. The first tool is RGA 7 from ReliaSoft [4] and the second one is **S**tatistical **M**odeling and **E**stimation of **R**eliability **F**unctions for **S**oftware (SMERFS).

## 2.2 Evaluation

Figure 1 presents the cumulative number of failures per time for the Skype application when applying the NHPP model. The RGA tool indicates an evident failure. Moreover, we tested the Vtok application and we found that the NHPP model also failed.

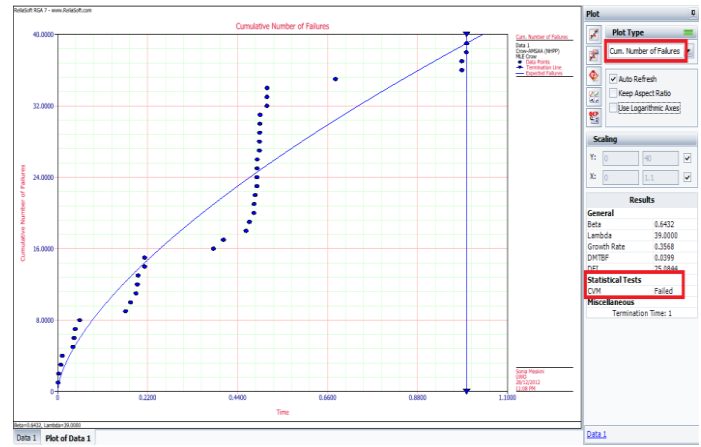

**Fig. 1.** Cumulative number of failures per Time (Skype).

In order to confirm our results, we used a second tool, SMERFS, and we applied the NHPP model to the same data points. The result was the same – the failure of the model each time. This failure can be traced back to the main differences between the desktop area and smartphones. One of the mobile application failure characteristics is that they are application dependent, in the sense that they are dynamic and non-homogenously spread in time. Moreover, they are unpredictable; sometimes they decrease and sometimes they increase. One possible explanation is that reliability depends on how the application is used, where it is used, and when it is used. The usage may differ from one person to another, from one country to another, from one condition and time to another, etc.; this explains the uncertainty of usage of the application in the execution and release time because all these factors play an important role in the reliability of the application.

Another reason is that the DLC (Development Life Cycle) of a mobile application is short (up to 90 days) and the programmer aims to develop the application as fast as possible to satisfy the time to market constraint, which leads to skip phases from the DLC. The phase most often skipped is the design phase, which is the most important phase in the DLC of the application [5]. Thus, it would be difficult to identify the causes of errors during the execution time and to find a convenient solution to fix them. Besides that, the failure or unreliability of the application may be caused by the technology used during the development process. The skills of the developer and the tester also play a huge role in the reliability of the application. Moreover, the device itself and its hardware characteristics – such as the size of the screen, the performance, the keyboard, etc. – can have a direct effect on the reliability of the application [6].

## 3 Failure Data Analysis Using Model Distributions

The preceding section was devoted to the application of the three most used SRGMs to two common smartphone applications, Skype and Vtok, and one private Windows phone application. The inputs to these models were the instantaneous failure data, i.e, the failure number and its exact time of occurrence. Those models failed to describe adequately the failure data. Having tried several non-linear models to better fit the failure data, we found that Weibull and Gamma distributions can be used to model new collected failure data of the same application after sorting them by version number and grouping them in different time periods [7]. Therefore, we used the two mentioned distributions and their particular cases, the Rayleigh and the S-Shaped models, and compared their performances for each application. This study was carried out in two steps: (1) the failure data for each application were sorted by version number and (2) the data were grouped by larger time scales (days, weeks, and months). An estimation of the total number of defects in each smartphone application version was obtained.

The Weibull distribution [8] is a two parameter function whose expression is given by:

$$f(t) = \text{wblpdf}(t,a,b) = \frac{b}{a} * \left(\frac{t}{a}\right)^{b-1} \exp\left(-\left(\frac{t}{a}\right)^b\right). \quad (1)$$

The parameters *a* and *b* take positive values as well as the variable *t*. If we define $A = 1/a^b$ and $B = b$, the expression simplifies to:

$$f(t) = B\,A\,t^{B-1}\exp(-A\,t^B). \quad (2)$$

A maximum for this function occurs at time $t = T_{max}$, such that

$$T_{max}^b = \frac{B-1}{A\,B}. \quad (3)$$

The Gamma distribution is a two parameter function whose expression is given by:

$$f(t) = \text{gampdf}(t,a,b) = \frac{1}{b^a \Gamma(a)}(t)^{a-1} \exp\left(-\frac{t}{b}\right). \quad (4)$$

for *a*, *b* and *t* taking positive values. The maximum of this function occurs at $t = T_{max}$, such that:

$$T_{max} = b(a-1). \quad (5)$$

### 3.1 Results

This section presents a comparison and an evaluation of the use of the above mentioned distributions to model the failure data of the Skype application, based on the usual evaluation criteria: RMSE, Ad-R-Square, and MRE. Due to space limitation, only Skype V1 will be presented from the versions we studied. The full and detailed results are found in [3].

For each application, the four distributions used were compared on the basis of their Root-Mean-Squared-Error (RMSE) and their Adjusted R-Square. The results of the estimated total number of defects were evaluated using the Magnitude of Relative Error (MRE). These statistical indicators are defined in [3].

Table 1 gives a compilation of all model parameters *(a, b)* along with the predicted or estimated $T_{max}$ (time of maximum failure rate) and the expected proportion $(Y(t = T_{max})/C)$ of encountered failures by $T_{max}$. It also gives results of RMSE, As-R-Square, the estimated cumulative number of failures *C*, and the MRE shown by each model. Only the best and the second best model distributions are given for each application version.

**Table 1.** Skype Version 1 – Error evaluation and model comparison.

| Skype V1 | Weibull | Gamma |
|---|---|---|
| Model parameters and deduced estimated values | a = 6.17 (5.26, 7.09)<br>b = 2.82 (1.81, 3.84)<br>Observed $T_{max}$ = 6<br>Estimated $T_{max}$ = 5.98<br>Estimated $(Y(t = T_{max})/C)$ = 47% | a = 6.14 (1.84, 10.44)<br>b = 0.97 (0.21, 1.73)<br>Observed $T_{max}$ = 6<br>Estimated $T_{max}$ = 5.01<br>Estimated $(Y(t = T_{max})/C)$ = 38.5% |
| RMSE | 2.1966 | 2.2305 |
| Ad-R-Square | 0.6374 | 0.6262 |
| C: Estimated cumulative number of failures or defects | 50.54 (34.51, 66.58) | 51.81 (34.32, 69.31) |
| MRE(%) | 6.4 | 4 |

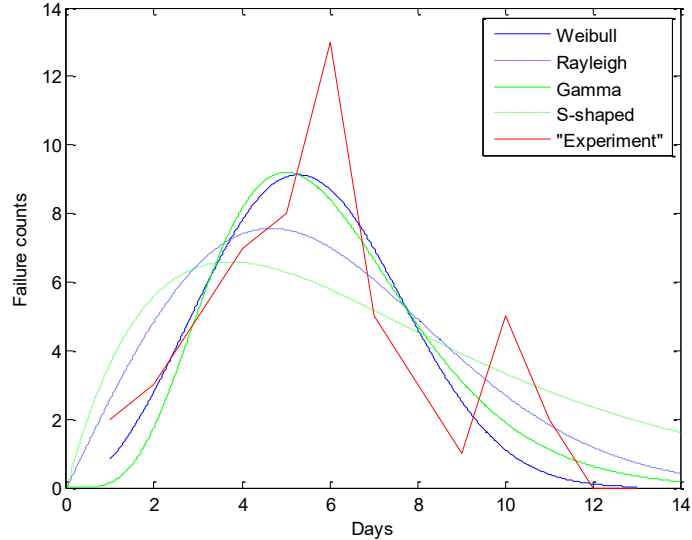

**Fig. 2.** Skype Version 1 – Model comparison.

Figure 2 portrays the results reported in Table 1. It can be noted from the figure that the Weibull distribution is the closest to the actual behavior curve of the application followed by the Gamma distribution.

## 4    Discussion and Answers to Research Questions

According to the preceding section, and as an answer to the first research question raised in the abstract, it can be concluded that the most successful reliability models [9] failed to account for the failure data and to predict the reliability of mobile applications. This failure can be traced back to the following main reasons: (1) Operational Environment and Usage Profiles of Smartphones Applications, (2) Hardware and Software Limitations [10].

Assuming all of these uncertainties, at a second stage, and in order to answer the second research question, we collected data from many users in different regions of the world, sorted them by application versions, and grouped them in different time periods (days, weeks, and months). Each application version failure data, when plotted in time periods, shows the same pattern: an early "burst of failures", due probably to the most evident defects, followed by a steep decrease in failure rate. After trying several non-linear models to fit the failure data, we found that the observed behavior is better modeled by the Weibull or Gamma distributions.

To answer the third research question the main features of this approach can be summarized as follows:
- For each application version, the model distributions are in fact distinguished by tiny differences in the calculated errors RMSE and Ad-R-Squared. Nevertheless, it can be concluded that no one single distribution can fit the data of all applications or even the different versions of the same application.

- As the parameters are given along with their 95% confidence intervals, it is to be noted that parameter *b* of the Weibull distribution, which is fixed to the value *b = 2* for the particular case of Rayleigh distribution, has confidence intervals that include the value *b = 2*. The same can be noted for parameter *a* of the Gamma distribution, which is fixed to the value *a = 2* in the particular case of the S-shaped model distribution. But most of the time, the general distribution models fit the failure data better than the particular cases.
- Similar to the famous 40% rule of the Rayleigh distribution, and independent of any application, the S-shaped distribution has a 26.4% rule. This means that by $T_{max}$, only 26.4% of the defects in a smartphone application will be uncovered. This can be tested on larger datasets and across many applications.

## 5   Conclusions

Our work is a step toward the application and evaluation of traditional Software Reliability models in the mobile area. We selected three of the most used models that are known for their efficiency in the desktop area: the NHPP, Musa-Basic, and Musa-Okumoto models. We examined two iPhone applications, Skype and Vtok, which were used and tested differently to evaluate the models under different conditions, and one Windows phone application. It turned out that none of the selected SRGMs was able to account for the failure data satisfactorily.

Our study also highlighted the causes of the failure of the models and the need for a meticulous SRGM for Smartphone applications, because the existing software reliability approaches were developed for traditional desktop software applications that are static and stable during their execution. This is not the case for smartphone applications, which have an unknown operational profile, a highly dynamic configuration, and changing execution conditions. On a continuous background, the smartphone failures come in relatively short bursts from time to time, which explain the abrupt in the observed cumulative failure number curves. This particular feature cannot be accommodated by the SRGMs that were used. Thus, in order to evaluate the reliability of smartphone applications, new models, principles, and tools are needed to incorporate the underlying uncertainties of such applications [11-14].

Our investigation of smartphone application reliability through the use of well-known available growth models suited primarily to desktop applications is twofold: (1) highlight the versatile nature of mobile applications, their dynamic configuration, unknown operational profile, and varying execution conditions in contrast to the static and stable desktop ones, and (2) stress the need for the design of new reliability models suited for mobile applications that take into account the inherent versatility of such applications [15]. Our future work will employ machine learning models [16-19] and will focus on analyzing these selected SRGMs in more depth and trying to modify the closest one to the data and adapt in to smartphone applications. Moreover, we will check to find out if we need to have a specific model for each type of applications or if one model is applicable to all the categories, of Smartphone applications, taking into consideration the severity of the failure.